# COVID-19 vaccination strategies on dynamic networks

Quoc Huy Nguyen[1], Jess Liebig[1,2], Md Shahzamal[3], Bernard Mans[3], Raja Jurdak[1]

[1]Trusted Networks Lab, School of Computer Science, Queensland University of Technology

[2] Health and Biosecurity, Commonwealth Scientific and Industrial Research Organisation

[3]Department of Computing, Macquarie University

## Abstract

Coronavirus disease (COVID-19), which was caused by SARS-CoV-2, has become a global public health concern. A great proportion of the world needs to be vaccinated in order to stop the rapid spread of the disease. In addition to prioritizing vulnerable sections of the population to receive the vaccine, an ideal degree-based vaccination strategy uses fined-grained contact networks to prioritize vaccine recipients. This strategy is costly and impractical due to the enormous amount of specific contact information needed. It also does not capture indirect fomite or aerosol-based transmission. We recently proposed a new vaccination strategy called Individual's Movement-based Vaccination (IMV) [5], which takes both direct and indirect transmission into account and is based on the types of places people visit. IMV was shown to be cost-efficient in the case of the influenza-like diseases. This paper studies the application of IMV to COVID-19 using its documented transmission parameters. We conduct large scale computer simulations based on a city-wide empirical mobility dataset to evaluate the performance and practicability of the strategy. Results show that the proposed strategy achieves nearly five times the efficiency of random vaccination and performs comparably to the degree-based strategy, while significantly reducing the data collection requirements.

## Introduction

The current COVID-19 outbreak not only affects global heath but has also caused major economic and social disruptions. Several vaccination campaigns are being implemented all over the world to minimize the overall impact. Previous research devised various theoretical vaccination strategies in both pre-emptive (before the disease widely spreads) and reactive (as the new virus strain generated) scenarios [1, 2]. The method of developing a vaccination strategy includes simulating the disease spread in a virtual environment with the selected strategy and interpreting the results. Various factors such as the number of vaccinated individuals and the cost of collecting contact information are taken into account to evaluate the performance of a vaccination strategy.

Existing vaccination strategies focus on local contact information of individuals such as where they have gone, how long they stayed in each location, or how many people they have met. One of the simplest contact-based vaccination strategies is randomly choosing a set of individuals in the network population to be vaccinated (RV) [2, 3]. Another simple strategy is acquaintance-based vaccination (AV) where individuals are asked to name a person they met [2, 3]. Individuals named by the majority will be vaccinated. However, these two strategies both require a large share of population to be vaccinated for it to be effective. To counter this problem, degree-based vaccination (DV) was proposed [4]. This strategy vaccinates individuals that have the highest number of contacts in a given period since they are the most likely disease spreaders. Even though this seems to be the most effective strategy, collecting the exact number of contacts for an individual is costly and difficult to implement. People tend not to remember how many people they have met and therefore, and often give inaccurate answers. While contact tracing apps can automate collection of this data, their uptake has generally been low due to privacy concerns.

We recently introduced a new strategy called individual's movement-based vaccination (IMV), that relies on the types of places a person visits rather than exact contact information [5]. In this strategy, individuals are ranked based on their frequencies of visits to different classes of locations. Locations are classified based on the potential number of contacts when an individual visits a certain place. Because it focuses on types of places visited rather than individual contacts, IMV improves privacy over the DV approach. Our work in [5] used the dataset collected from a location-based app over 32 days and performed traced-based simulations to evaluate the performance of each mentioned strategy for influenza-like illnesses. Although DV has the best performance in both pre-emptive and reactive scenarios, IMV achieves comparable performance while being realistically implementable [5].

In this paper, we comparatively evaluate the vaccination strategies for COVID-19 and explore the impact of emerging and more infectious COVID-19 strains on the efficiency of these strategies.

## Materials

### Individual's movement based vaccination

The individual's movement based vaccination ranks each individual based on movement behaviors instead of exact measure of contact information. Each individual is asked about the frequencies of visiting different classes of locations, and then, is given a ranking score, indicating their impact in spreading the disease. Locations such as public transport hubs, offices or shopping malls are classified based on the estimations of the how many contacts could occur during one visit. Table 1 shows six classes of locations. We describe the details of assessing an individual's rank below, as per our approach in [5].

Let $\beta$ be the probability that a susceptible individual $v$ who visits location $L$, where an infected individual $u$ has visited, gets infected. If an infected individual $u$ meets $d$ individuals during this visit, the probability of transmitting the disease to any contacts through this visit is given by [5]:

$$w = 1 - (1 - \beta)^d$$

Here, the assumption is that the individual $u$ is only the source of infection. All contacts are susceptible and in contact with $u$ independently. Under these assumptions, let $w$ be the spreading potential for a visit by an infected individual at a location with a number of individuals. The spreading potential for visiting a location belonging to a class $i$ can be approximated as:

$$w_i = \frac{1}{2}\left(2 - (1-\beta)^{d_i^1} - (1-\beta)^{d_i^2}\right)$$

where $d_i^1$ is the lower limit of class $i$ and $d_i^2$ is the upper limit of class $i$. In fact, this is the average spreading potential for the class $i$ locations. The ranking score of an individual for visits to different classes of locations is given as:

$$W = \sum_1^6 f_i \times w_i$$

where $f_i$ is the frequency of visits to a location belonging to class $i$.

The individual ranking score $W$ represents the maximum number of disease transmission events during the observation period. As this score is a relative value, it will carry significant information even if the same neighbouring individuals meet repeatedly. This is because repeated interaction increases the disease transmission opportunity. In addition, $W$ indicates how easily a susceptible individual $v$ gets infected due to movement behaviours and the frequency of interactions. Like the degree-based strategy (DV), the IMV strategy accounts for super-spreaders, but also accounts for the intensity of interactions among individuals through $d_i$ and $f_i$. It also accounts for the importance of places that DV strategy does not consider, such as locations where indirect transmission is likely due to fomites or aerosols. In practice, it is also easier to remember the visited locations than how many people one has met.

Table 1: Classification of visits nodes do during their daily activities.

| class | contact sizes | locations |
|---|---|---|
| class-1 | 1-5 | home, store |
| class-2 | 6-15 | coffee shop, bus stop |
| class-3 | 16-25 | office, local train station, small park, swimming pools |
| class-4 | 26-50 | central train station, large park, clubs and night clubs |
| class-5 | 51-100 | shopping mall, college, central train station, sport centers |
| class-6 | 101- | university, college, airport, concerts, festival and football stadiums |

https://doi.org/10.1371/journal.pone.0241612.t001

**Dataset**

This study conducts experiments on contact information of a location-based app called Momo. The app provides location updates every time users launch it. This data contain millions of records from Momo users all around the world, each of which includes Earth coordinates, time of the update and user identifications. An SPDT (same place different time) contact network in 32 days (from 17 September, 2012 to 19 October, 2012) was extracted from the data by applying the SPDT diffusion model [6, 7]. However, the SPDT network is sparse, meaning the majority of the nodes have small degrees, which can lead to underestimation of an individual's likelihood to spread the disease. Hence, a dense SPDT network was constructed (DDT), by randomly duplicating contact links of users [5]. The final DDT network contains millions of contact links for over 360,000 users.

**Disease Propagation**

To simulate the spread of the disease on the DDT network, we used the formula about the probability of infection for each interaction defined as

$$P_I = 1 - e^{-\sigma E} \quad (1)$$

where $\sigma$ is the infectiousness of the disease and $E$ is the exposure for contacting with the infected node [5]. If a susceptible individual got infected with the probability $P_I$, they enter the latent period (time from infection to being able to transmit disease) in day $\alpha$, then continue to produce infectious particles to day $\beta$, before they enter the recovered state.

The disease was simulated in a pre-emptive scenario where one node is randomly chosen to be infected. Each link in the network is considered separately with a particular infection probability drawn from Equation (1). The incubation period (time from infection to symptom onset) $x$ is generated with the log-normal distribution with mean of 1.621 and standard deviation of 0.418 [8]. The latent period is approximately three days shorter than the incubation period [9] and therefore, $\alpha$ is set to the day individual got infected plus $x - 3$. The infectious period is set to 11 days, beginning three days prior to the onset of symptoms [9, 10], making $\beta$ to be $\alpha + 11$. The infectiousness $\sigma$ is interpreted to the R value of COVID-19 (the number of infections caused during the infectious period of a single infective). The 2020 statistic recorded the R value of coronavirus to be most likely close to 1 [11]. Therefore, the infectiousness $\sigma$ is first set to 1, then, later changed according to the new R value recorded for new COVID-19 strains.

**Pre-emptive simulations**

The purpose of pre-emptive vaccination is to prevent or hinder the future spread of the disease. We conducted simulation without any vaccination to record the upper bound of the number of infections. No lockdown, quarantine, or any protocols to reduce the impact of the virus was taken into account. Then, a proportion of the population chosen by each strategy is vaccinated.

Various vaccination rates P (percentage of *N* total individuals) are chosen, meaning *PN*/100 nodes are vaccinated by ranking each individual with mentioned strategies using data from the first seven days. Then, a random seed node is chosen to be infected and obtain the outbreak size (number of infections) after propagating the disease for 25 days. To account for simulation randomness, the process is repeated 1000 times, each with a different seed node, and the average outbreak size is recorded.

**Results**

The average outbreak size without any vaccination for DDT network is 3365 infections, or roughly 1% of the total population, providing a baseline scenario. Figure 1 shows the outbreak size for each of the four vaccination strategies as the vaccination rate P varies. The efficiency of each strategy with an outbreak size S is calculates as (3365-S)/3365. It clearly shows that RV has little impact on limiting the disease spread with the highest efficiency of only 19.0% for a vaccination rate $P = 1.8\%$. The AV strategy does improve the efficiency up to around 82.2%. The IMV and DV have comparable performance with the highest efficiency of 99.0% for IMV and 99.6% for DV both on 2.0% vaccination rate.

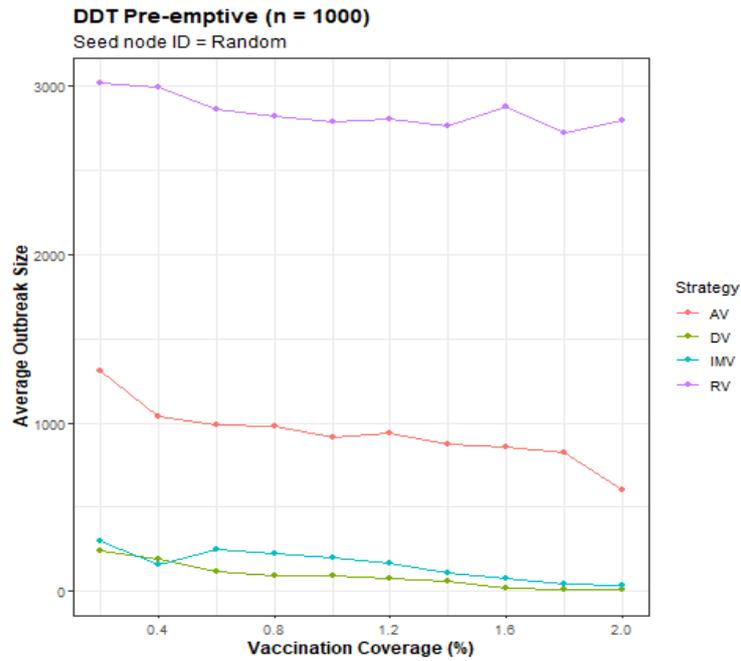

**Figure 1: Average outbreak size versus vaccination coverage rate with pre-emptive vaccination**

Next, we explore the required vaccination rate P that results in a contained outbreak size within 100 infections, shown in Figure 2. The IMV only requires 1.6% vaccination rate to get under 100 infections while RV needs at least 86.0% and AV needs 65.0%. At a low vaccination rate $P$ (less than 1.0%), the DV strategy outperforms IMV but the gap gets smaller as $P$ increases.

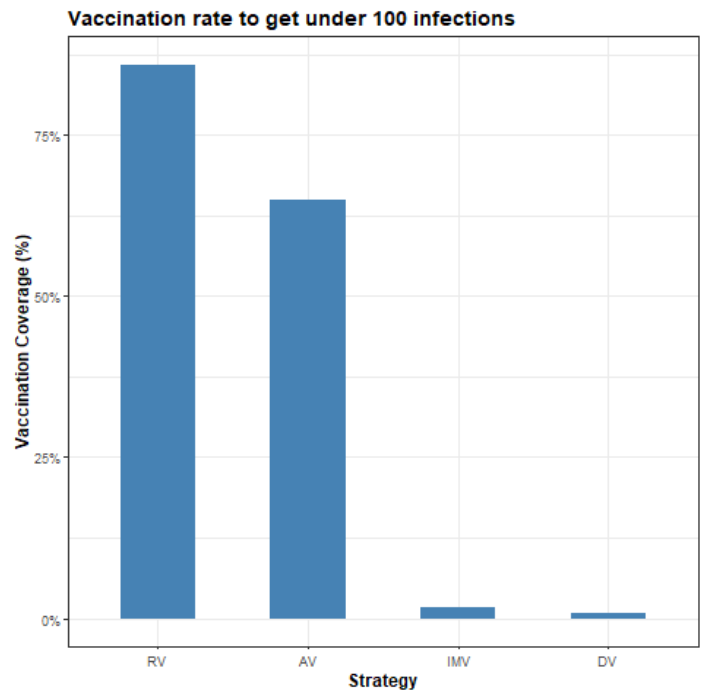

**Figure 2: Vaccination rate to get under 100 infections**

**Different infectiousness profiles**

The R value used in previous simulations was recorded in October 4, 2020 [11]. However, R may vary across regions or countries due to various factors such as lockdown protocols, environmental conditions, or new virus strains. Recent reports from UK presents a new variant of coronavirus resulting in a 70% increase in the R value [12]. This section repeats the simulations but with different values of R.

The experiments are conducted to evaluate the efficiency of the strategies varying the infectiousness $\sigma$ in Equation (1). The vaccination rate remains constant at 1.0% and R value varies from 1.0 to 1.2 and 1.7.

The results in Figure 3 show that the IMV and DV strategies remain efficient for all variants of the R value. With the R value increased to 1.7, no vaccination strategy is able to keep the number of infections under 100. Among the realistic vaccination strategies, IMV reduces outbreak size of RV and AV by a factor around 8 and 6 respectively for the most infectious scenario with an R-value of 1.7.

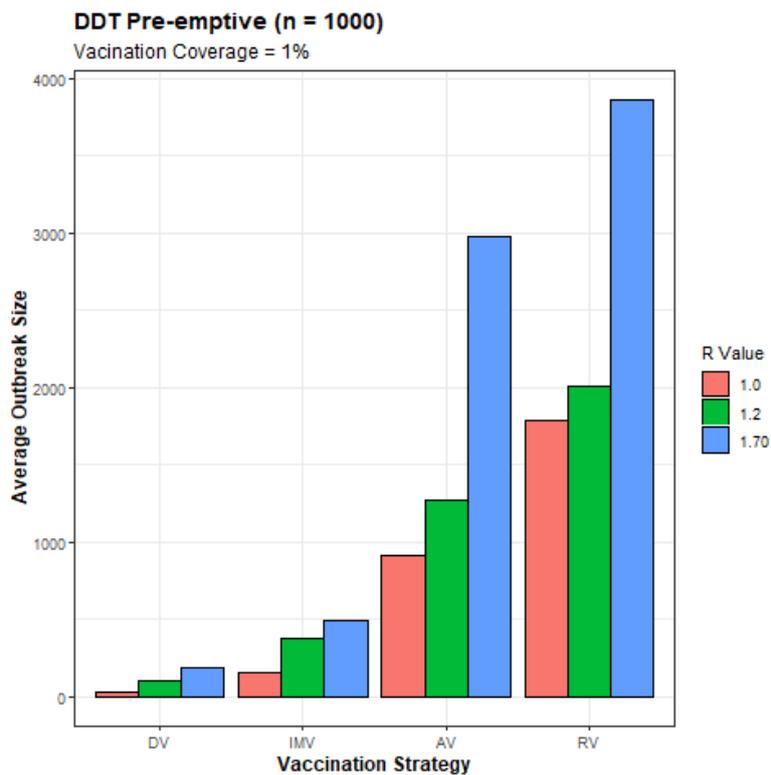

Figure 3: Different R value with 1% vaccination rate

# Conclusion

This paper has comparatively evaluated vaccination strategies for COVID-19 on dynamic networks of people movement. Using empirical location traces, we found that the visitation place-based IMV strategy achieved a five-fold improvement over random vaccination and comparable performance in vaccination efficiency to the idealized degree-based strategy. Its benefit lies in its cost efficiency, as it requires only the types of places people visit and the frequency of visits. Our simulations exploring the effect of R-value reveal sustained benefits of IMV over the other practical strategies. Our findings can help inform ongoing rollouts of COVID-19 vaccines to maximise their efficiency in containing outbreaks.